\documentclass[pra,showpacs,aps,amssymb,amsmath,nofootinbib,notitlepage,superscriptaddress,twocolumn,longbibliography]{revtex4-2}

\usepackage{graphicx,graphics,epsfig,times,bm,mathrsfs}
\usepackage[normalem]{ulem}
\usepackage{subfigure}
\usepackage{xfrac}
\usepackage{mathtools}
\usepackage{amsmath,amssymb,amsfonts}
\usepackage[table, xcdraw]{xcolor}
\usepackage{varwidth}
\usepackage{babel}
\usepackage{hhline}
\usepackage{multirow}
\usepackage{tikz}
\usepackage{longtable}
\usepackage{pgf}
\usepackage{physics}
\usepackage{bbm}
\usepackage{paralist}
\usepackage[hidelinks]{hyperref}
\usepackage{float}
\usepackage{microtype}
\usepackage[T1]{fontenc}
\usepackage[utf8]{inputenc}
\usepackage{tikzsymbols}
\usepackage{algorithm}
\usepackage[noend]{algpseudocode}

\tikzset{>=latex}
\usetikzlibrary{
    external,
    patterns,
    shapes.geometric,
    arrows,
    positioning,
    chains,
    decorations.pathmorphing,
    decorations.markings,
    decorations.pathreplacing,
    shapes,
    shadows.blur,
    shapes.symbols,
    calligraphy,
    fadings,
    calc,
}
\pgfdeclarelayer{bg}    
\pgfsetlayers{bg, main}  

\newcommand{\enquote}[1]{``#1''}

\begin{document}

\title{Quantum key distribution in a packet-switched network}

\author{Reem Mandil}
\email{reem.mandil@mail.utoronto.ca} 
\affiliation{Cisco Quantum Lab, Los Angeles, California, USA}
\affiliation{University of Toronto, Toronto, Canada}

\author{Stephen DiAdamo}
\email{sdiadamo@cisco.com}
\affiliation{Cisco Quantum Lab, Garching bei M\"unchen, Germany}

\author{Bing Qi}
\email{bingq@cisco.com}
\affiliation{Cisco Quantum Lab, Los Angeles, California, USA}

\author{Alireza Shabani}
\affiliation{Cisco Quantum Lab, Los Angeles, California, USA}

\date{\today}

\begin{abstract}
Packet switching revolutionized the Internet by allowing the efficient use of network resources for data transmission. In a previous work, we introduced packet switching in quantum networks as a path to the Quantum Internet and presented a proof-of-concept for its application to quantum key distribution (QKD). In this paper, we outline a three-step approach for key rate optimization in a packet-switched network. Our simulated results show that practical key rates may be achieved in a sixteen-user network with no optical storage capacity. Under certain network conditions, we may improve the key rate by using an ultra-low-loss fiber delay line to store packets during network delays. We also find that implementing cut-off storage times in a strategy analogous to real-time selection in free-space QKD can significantly enhance performance. Our work demonstrates that packet switching is imminently suitable as a platform for QKD, an important step towards developing large-scale and integrated quantum networks.
\end{abstract}

\maketitle

\section{Introduction}\label{sec:intro}
\begin{figure} 
    \begin{center}
        \subfigure[]{\includegraphics{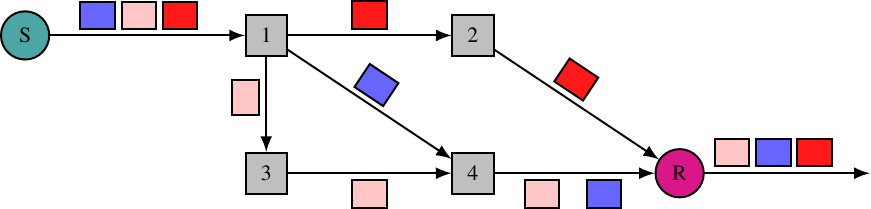}} \label{fig:packet-switching}
        \subfigure[]{\includegraphics{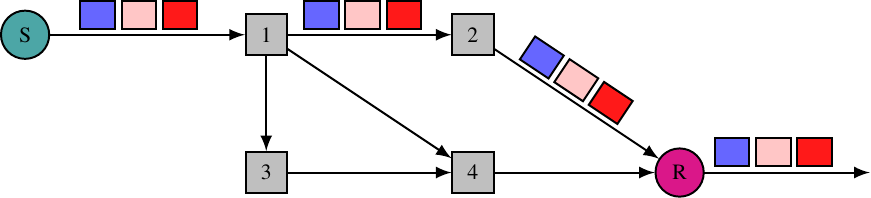} \label{fig:circuit-switching}}
    \end{center}
    \caption{(a) Packet-switched network. The channel between sender (S) and receiver (R) is not predetermined and can be dynamically reconfigured. (b) Circuit-switched network. A dedicated channel between sender and receiver is set up before data is transferred between them.}
\end{figure}

Packet-switched communication networks were introduced as an efficient and scalable alternative to circuit switching in the early sixties~\cite{kleinrock, baran1964}. Today, packet switching is the dominant mode of operation in the Internet. Recently we have introduced packet switching as a paradigm for quantum networks using hybrid (classical-quantum) data frames~\cite{diadamo2022packet}. Inside a frame, a quantum payload is prepended with a classical header containing information for routing and more. Frames travel from sender to receiver through a series of routers which process the header to determine the channel forward based on the current conditions of the network (Fig.~\ref{fig:packet-switching}). This is in contrast to a circuit-switched network where a dedicated channel is established between sender and receiver and reserved until communication is complete (Fig.~\ref{fig:circuit-switching}). 

There are important considerations to be made when deciding whether packet switching or circuit switching is best suited for a network application. In a circuit-switched network, communication across multiple user pairs must be done in a coordinated fashion in order to enable bandwidth sharing (e.g., via time or wavelength-division multiplexing). In a packet-switched network, the communication need not be coordinated in advance. However, frames will experience delays at the intermediate nodes between users due to finite header processing times and, under some traffic conditions, queuing times. Furthermore, packet switching is generally advantageous over circuit switching when the traffic generated by network users is \textit{bursty}, characterized by intervals of activity and intervals of inactivity. 

One important application in a quantum network is quantum key distribution (QKD), a procedure that allows two remote users (e.g., Alice and Bob) to establish shared encryption keys with information-theoretic security~\cite{bb84, e91}. An important feature of QKD is that it is robust against loss in transmission, meaning that a secure key can still be established even when most of the transmitted signals are lost. This suggests that data loss due to delays in a packet-switched network may be tolerated even without any storage of QKD signals at the routers. Moreover, the optical loss introduced by an imperfect storage medium may also be tolerated. Another important feature of QKD is that key generation is not time-critical, meaning that secure keys need not be generated immediately before their consumption. This implies that bursty frame generation may be sufficient since users can establish and store keys for later use. 

These features motivate our hypothesis that packet switching is imminently suitable as a platform for QKD. One may of course imagine a scenario where network users prefer access to a dedicated quantum channel for their key distribution (e.g., urgent requests or large size requirement for encryption keys). Furthermore, most existing demonstrations of multi-user QKD are conducted over dedicated networks~\cite{Townsend1997,Elliot2005,Peev2009,Sasaki2011,Frohlich2013,Tang2016,Chen2021} where QKD is the sole task. In this case, it may be beneficial to have a central controller to coordinate QKD among different user pairs, in a fashion similar to circuit switching. However, if we wish to integrate QKD with existing classical networks in order to extend its applications, packet switching is a natural choice. Therefore, the goal of this paper is to demonstrate the feasibility of performing QKD in a packet-switched network. 

To meet this goal, we take a three-step approach. First, we choose a network routing protocol which describes how a router handles a frame during network delays. In this paper, we will investigate three different routing protocols based on varying optical storage capacity. Second, we simulate the transport of frames in a network operating under a given routing protocol and traffic model. The simulation provides us with statistics for the dynamic channel between each Alice-Bob pair. Lastly, we use the simulated network statistics to predict the maximum secure key rate for each user pair in the network by performing a finite-key analysis.      

In our previous work~\cite{diadamo2022packet}, we presented a proof-of-concept for QKD in a packet-switched quantum network, and considered a basic model for a two-user communication scenario where the routers had no optical storage capacity. Packet switching in quantum networks is a relatively unexplored topic, but has been proposed as a solution for overcoming scalability issues in previous works~\cite{munro2022designing, yoo2021quantum}. Moreover, Ref.~\cite{singal2022hardware} has investigated using leading classical signals to make routing decisions in a QKD network, although packet switching is not considered in their approach. In this work, we analyze a sixteen-user network with and without optical storage capacity at the routers. We also consider a finite-size security analysis for a practical decoy-state QKD protocol. Our results show that QKD is feasible in a packet-switched network with today's commercial technology and that optical storage can be used to improve its performance under certain conditions. 

This paper is organized as follows. In Sec.~\ref{sec:routing}, we describe the routing component of a packet-switched network, including network delays and the routing protocols considered in this work. We also present a router hardware design based on current technology. In Sec.~\ref{sec:security}, we describe the QKD protocol and key rate analysis under consideration. In Sec.~\ref{sec:simulation}, we describe our software tool for simulating the dynamics of a packet-switched network. Finally, in Sec.~\ref{sec:results}, we present and discuss the simulated QKD results. 

\section{Network Routing}\label{sec:routing} 
In this section, we describe how the routers in a packet-switched network may handle frames that are intended for a QKD application. We review the frame structure and outline the network delays and routing strategies considered.

\subsection{Network Delays}
The total time a frame needs to move through a router is the sum of three sources of delay. First, there is the processing delay, $d_{proc}$, which is the time to process the classical header and determine the next action for the frame as well as regenerate the header when needed. Depending on the network complexity, this delay can range from 10~$\mu s$ to 1,000~$\mu s$~\cite{ramaswamy2004}. In this work, we assume a $d_{proc}$ of 100~$\mu s$. Second, there is the queuing delay, $d_{queue}$, which is the time the frame must wait before it can be forwarded from a router (after the header has been processed). This quantity depends on the traffic conditions of the network and can range from zero to infinity. Lastly, there is the transmission delay, $d_{trans}$, which is the time required to transmit the entire frame onto an outgoing link. This is equal to the temporal frame length, $T_f$, which may shrink at each router it traverses depending on the routing protocol employed. 

\subsection{Routing Protocols}\label{subsec:protocols}

\begin{figure} 
    \begin{center}
        \subfigure[\label{fig:no_guard}]{\includegraphics{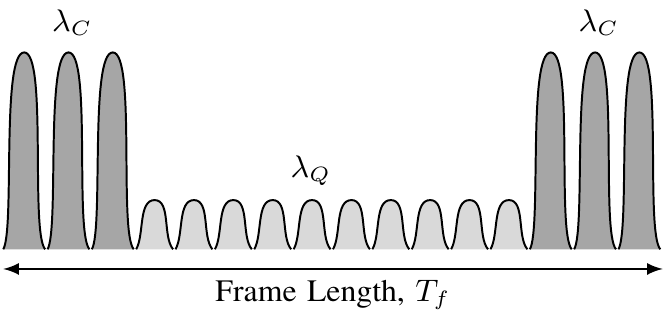}}
        \subfigure[\label{fig:guard}]{\includegraphics{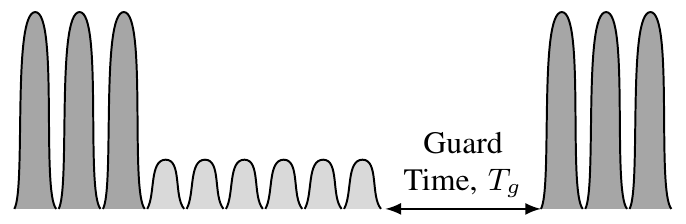}}
    \end{center}
    \caption{(a) The classical header and trailer ($\lambda_C$) and the quantum payload ($\lambda_Q$) are generated from a laser source and multiplexed into a hybrid data frame using time-division and wavelength-division multiplexing (not shown to scale). (b) The hybrid frame includes guard time---a time delay between the end of the header and the beginning of the payload.}
    \label{fig:frame}
\end{figure}

The network routing protocol determines what happens to a frame during the network delays $d_{proc}^i$ and $d_{queue}^i$, where the superscript $i$ is used to index each router in the frame's path from sender to receiver. Fig.~\ref{fig:frame} depicts a hybrid frame with a quantum payload consisting of weak laser pulses with repetition rate $R_t$ (Hz). The frame may be configured to include a time delay between the end of the header and the beginning of the payload, referred to as the guard time, $T_g$. 

In general, our network routing protocols fall into one of two categories based on the capacity to store frames at the routers. For protocols based on no storage, $d_{trans}^i$ ($= T_f^i$) will shrink by a duration equal to $d_{proc}^i+d_{queue}^i$ at each router the frame traverses. If $T_g^i = 0$, this corresponds to the discarding of $R_t(d_{proc}^i+d_{queue}^i)$ pulses in the leading portion of the payload (note that we consider the lengths of the classical header and trailer to be negligible compared to the quantum payload). If $T_g^i > 0$, then it will serve as a buffer to reduce the number of pulses that are lost (i.e., if $T_g^i > d_{proc}^i+d_{queue}^i$, then no pulses are discarded as the frame shrinks but $T_g^i$ will decrease accordingly). Note that in each routing protocol we consider, the guard time is not reset at each router. This alternative approach may be useful for a quantum network application in which the payload carries information that should not be lost.   

For protocols based on storage, the frame will enter a fiber delay line for a storage time $T_s^i \leq d_{proc}^i+d_{queue}^i$. During $T_s^i$, no pulses are discarded from the payload, but they will be subject to the attenuation of the fiber delay line. If $T_g^i > 0$, then it will again serve as a buffer to reduce $T_s^i$ (i.e., if $T_g^i > d_{proc}^i+d_{queue}^i$, then $T_s^i=0$ but $T_g^i$ will decrease accordingly). Note that the header may be configured to include a field that tracks the cumulative time spent in storage as a frame traverses the network. In this work, we investigate the following three routing protocols.
\begin{enumerate}
    \item \textit{No storage during delays.} At each router, a frame will have its payload discarded for a time $d_{proc}^i+d_{queue}^i$ and $d_{trans}^i$ will shrink by the same amount. If $d_{trans}^i$ reaches zero, then the frame is discarded from the network.
    \item \textit{Storage during delays (unlimited).} At each router, a frame will enter a fiber delay line for a storage time $T_s^i = \max(0, d_{proc}^i+d_{queue}^i-T_g^i)$ and $d_{trans}^i$ will shrink by $\min(T_g^i, d_{proc}^i+d_{queue}^i)$.
    \item \textit{Storage during delays (limited).} At each router, a frame will enter a fiber delay line for a storage time $T_s^i = \max(0, d_{proc}^i+d_{queue}^i-T_g^i)$ and $d_{trans}^i$ will shrink by $\min(T_g^i, d_{proc}^i+d_{queue}^i)$. If the total time a frame has spent in storage reaches a predetermined storage time limit, the frame is immediately discarded from the network. 
\end{enumerate}

In the no storage routing protocol, network delays introduce a controlled photon loss as a portion of the payload is discarded. In the storage routing protocols, network delays introduce random photon loss in the payload due to the attenuation of the fiber delay line. The regime in which one strategy may dominate over the other therefore depends on factors such as the frame length, the network delays, and the attenuation of the storage line. A more detailed motivation for the two types of routing protocols is provided in Appendix~\ref{sec:nostorage_vs_storage}.   

To motivate the limited storage routing protocol, we make the observation that the dynamic channel conditions in a packet-switched network are analogous to those in free-space QKD under turbulent conditions. In such scenarios, it has been shown that the key rate can be improved by rejecting key bits when the channel's transmittance is below a threshold~\cite{Erven2012, Vallone2015, Wang2018PRTS}. In our case, since the routing history is recorded in the classical header, we can discard frames en-route, which has the additional benefit of reducing network congestion. Another option, more analogous to the technique used in free-space QKD, is to allow all frames to reach the receiver end via the unlimited storage routing protocol, but enforce a storage time limit (STL) in post-processing. That is, frames for which $\sum_i T_s^i > STL$ will be excluded from key generation. In this work, we compare both options for implementing a cut-off channel transmittance. 

\subsection{Router Hardware}

\begin{figure*} 
    \centering    
    \begin{center}
        \includegraphics{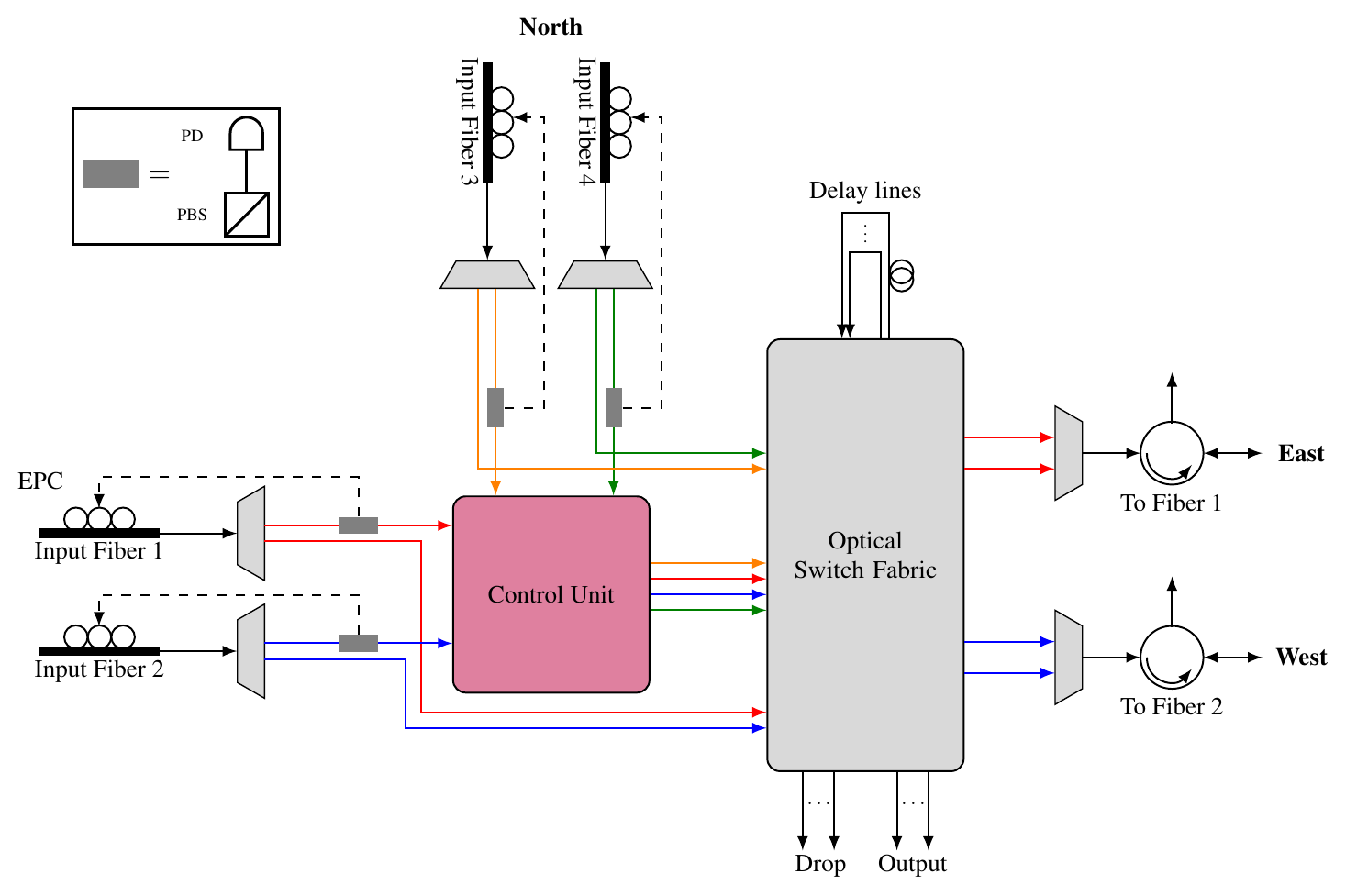}
    \end{center}
    \caption{Hardware design of a router in a packet-switched network. A frame arrives at the router from the North, East, or West degree. Channels in the North degree are directly connected to senders. The links in the East and West degrees consist of a fiber directly connected to another router; a circulator is used to allow for bidirectional transmission. A frame passes through a wavelength-division multiplexer to separate the classical and quantum information. The classical information is processed in the control unit, which signals to the optical switch fabric where to route the frame (i.e., to another router via the East or West degree, or to a receiver via an Output channel) and regenerates the header prior to transmitting the frame. Add/Drop channels are used to access variable optical fiber delay lines. Drop channels are used for discarding pulses or entire frames. PD: photodiode; PBS: polarizing beam splitter; EPC: electronic polarization controller.}
    \label{fig:router}
\end{figure*}

A conceptual router design is shown in Fig.~\ref{fig:router}. This router behaves as a quantum version of a reconfigurable optical add drop multiplexer (ROADM). Frames may arrive at the router from three different directions (North, East, West) after which a wavelength-division multiplexer is used to separate the quantum payload from the classical header and trailer. The header is fed into a control unit to decide how to further process the frame. Once the header has been processed, the frame will be forwarded towards the next node in the network (i.e., to another router via the East or West degree, or to a receiver via an Output channel). The control unit will regenerate the header with updated fields for the quantum payload duration, guard time, and time spent in storage prior to transmitting the frame to the next node. 

We assume the control unit is capable of processing up to $k$ headers simultaneously and that the router has access to $q$ variable optical fiber delay lines via its Add/Drop channels. To achieve an arbitrary delay, each fiber delay line can be combined with an active optical switch (not illustrated in figure). The router can also discard frames or partially discard the quantum payload via its Drop channels. The use of these channels depends on the network routing protocol being implemented. 

We also assume the router to have a minimum insertion loss of 4 dB, which accounts for the circulators, multiplexers, and optical switch fabric (excludes the fiber delay lines). Therefore the total loss (dB) at each router is given by
\begin{equation}\label{eq:router_loss}
    loss_r^i = T_s^i v_g \alpha_s + 4 \hspace{1mm} dB,
\end{equation} where $v_g$ is the speed of light in fiber and $\alpha_s$ is the attenuation coefficient (dB/km) of the fiber storage line. Furthermore, we assume the router may compensate the polarization drift of all incoming channels by using a feedback signal generated from the measured drift of the classical pulses in the header. 

Lastly, we note that this router design is directly suitable for the network configuration in Fig.~\ref{fig:network-topology} although additional input fibers and ROADM degrees may be added to the router depending on the desired connectivity of the network. We also consider hardware that is directly suitable for the hybrid frame in Fig.~\ref{fig:frame} although the hardware can be modified according to the multiplexing scheme employed for the frame.

\section{QKD Security Analysis}\label{sec:security}
Practical implementations of QKD adopt the decoy-state method~\cite{hwang2003quantum,lo2005decoy,ma2005practical,wang2005beating} to allow for use of a weak pulsed laser source instead of an ideal single-photon source. In this work, we consider a decoy-state asymmetric coding BB84 protocol~\cite{lo2005efficient} and we adopt the finite-size security analysis in Ref.~\cite{lim2014concise} to calculate the secure key rate. In this section, we provide a brief summary of the QKD protocol and then describe our strategy for key rate optimization in a packet-switched network. 

\subsection{Protocol Description}\label{subsec:QKDprotocol}
\textit{1. Preparation.} Alice chooses a bit value $b_A$ uniformly at random. Then, she selects a basis $\in \{X,Z\}$ with probabilities $q_x$ and $1-q_x$, respectively, and an intensity $k_i \in \mathcal{K} \coloneqq \{\mu_1, \mu_2, \mu_3\}$ with probabilities $p_{\mu_1}$, $p_{\mu_2}$, and $p_{\mu_3} = 1-p_{\mu_1}-p_{\mu_2}$, respectively. If Alice chooses the $X$ basis, she prepares a weak laser pulse of the chosen intensity in the horizontal polarization state $\ket{H}$ for the bit value $b_A=0$ or vertical state $\ket{V}$ for the bit value $b_A=1$. If the $Z$ basis is chosen, she prepares the diagonal (45-degrees) polarization state $\ket{D}$ for the bit value $b_A=0$ or antidiagonal (135-degrees) state $\ket{A}$ for the bit value $b_A=1$. Lastly, she sends her prepared state to Bob.

\textit{2. Measurement.} Bob selects a basis $\in \{X,Z\}$ with probabilities $q_x$ and $1-q_x$, respectively. Then, he performs a measurement in the chosen basis and records the outcome in a bit value $b_B$. More precisely, he assigns $b_B=0$ for a click in single-photon detector $D_0$ and $b_B=1$ for a click in detector $D_1$. If both detectors click, he assigns a random value to $b_B$. If neither detector clicks, he does not assign any value.  

\textit{3. Basis reconciliation.} Alice and Bob announce their basis and intensity choices over an authenticated public channel. Based on the information announced, Alice and Bob identify their raw keys $b_A$ and $b_B$ from the instances where they both chose basis $X$ and Bob observed a detection event. Note that all intensity levels are used for the key generation~\cite{lim2014concise}. They use the instances where they both chose basis $Z$ and Bob observed a detection event for phase error estimation.  

\textit{4. Post-processing.} Alice and Bob perform classical error correction and privacy amplification on their raw key pair to extract a secure key. 

\subsection{Key Rate Optimization}
A convenient feature of QKD security proofs is that the quantum channel between users is assumed to be fully controlled by an adversary and thus we do not need to develop a new security proof for QKD in a packet-switched network. One may ask whether we need to trust the routers which control the discarding of pulses and frames. If a security proof allows for the adversary to fully control Bob's post-selection process, as is the case for the proof adopted in this work, then we need not trust the routers. Nonetheless, packet switching poses a unique challenge to QKD due to the dynamic nature of the quantum channel between users. In order to maximize the secure key rate in the decoy-state protocol described above, we must optimize over the free parameters $\{q_x, p_{\mu_1}, p_{\mu_2}, \mu_1, \mu_2\}$~\cite{lim2014concise} which requires knowledge of the average channel transmittance, $\langle \eta_{tot} \rangle$, where the average is taken over all frames contributing to the key. Furthermore, in order to conduct a finite-size analysis, we must determine the total number of QKD states, $N$, passed to Bob. Depending on the network routing protocol employed, this may not be equivalent to the number of states transmitted by Alice, $N_0$, due to discarding at the routers. Therefore, in order to predict the maximum secure key rates from QKD in a packet-switched network, we need a tool for assessing $\langle \eta_{tot} \rangle$ and $N$ for each user pair. One may consider an analytic approach to gathering these statistics, however this quickly becomes infeasible for increased complexity of the network. The theory of Jackson networks~\cite{jackson1957networks} allows us to calculate the average queuing delay at each router quite simply, but only if the network obeys a specific traffic model. Instead, we build a network simulation tool to numerically determine the channel statistics. Details of the key rate analysis, including noise and detection parameters, are given in Appendix~\ref{sec:r_calc}. 

\section{Network Simulation}\label{sec:simulation}
In this section, we first provide a high-level description for the sequence of events that occur as a frame travels from sender to receiver in a packet-switched network and then describe our software tool for simulating these events in order to extract the dynamic channel statistics.

We model the arrival of frames into the network as follows. Each sender is allowed to transmit frames one at a time, following an exponentially distributed inter-arrival time with average $1/\gamma$. Note that all senders can be active simultaneously. We assume a repetition rate $R_t = 1$~GHz for the signals in the quantum payload. The destination for each frame is assigned randomly from the list of all receivers in the network. 

A frame travels from a sender towards its default router (i.e., the router to which the sender is directly connected). The default router and all subsequent routers a frame encounters will forward the frame according to the path determined by the routing algorithm for the network. The routing algorithm calculates the least-cost path from sender to receiver, where the cost of a path is the sum of the link costs along the path. In this work, we consider a load-insensitive routing algorithm, meaning the cost of each link in the network does not reflect its level of congestion and is determined solely by its physical length. Therefore, the least-cost path is simply the shortest path. Note that in the case of multiple least-cost paths, the router will select one at random. In general, the shortest path may not have the highest expected transmittance, depending on the number of routers it contains. In this case, the cost of the path may be modified to include router loss, although this scenario is not applicable in this work. 

A frame can be forwarded from a router only if there are fewer than $c$ frames simultaneously being forwarded from the router and there are no frames preceding it in the queue (we refer to $c$ as the number of servers for the queue); otherwise, the frame must join the queue. A frame may join the queue only if there are fewer than $q$ frames already in the queue (we refer to $q$ as the capacity of the queue); otherwise, the frame will be discarded. Frames will be forwarded from the queue according to a first-come first-served discipline.

In order to simulate these events in a network, we developed a software tool based on a simulation method known as discrete-event simulation (DES)~\cite{matloff2008introduction}. We build on the DES Python package \textit{SimPy}~\cite{muller2003simpy} for the timing and resource management aspects of the network. For the network configuration, including path calculations and topology initialization, we use the Python package \textit{NetworkX}~\cite{hagberg2008exploring}.

The first step in using our simulation is to configure a topology of nodes (i.e., users and routers) and links (i.e., connections between nodes). Each node is able to generate frames as well as process any incoming frames. If the node is a sender, frames at the node do not undergo header processing and the frame need only wait to be sent into the network according to the frame arrival model. If the node is a router or a receiver, frames at the node will undergo a processing delay. In our simulation, routers can process $k\gg 1$ headers simultaneously. In general, if $k$ is small, the frames may experience a queuing delay prior to header processing. In our simulation, the queue in each router has $c=1$ server and unlimited storage capacity ($q\rightarrow \infty$). The actions on the frame during the processing and queuing delays will depend on the network routing protocol, as outlined in Sec.~\ref{subsec:protocols}. 

Each frame in the network holds attributes (corresponding to header fields) for the storage time limit, how long it has spent in storage, the temporal frame length, the guard time, the path it has travelled, and its status (in transit, arrived, or discarded). We can simulate the network dynamics for a specified duration and collect data on the number of routed QKD signals, $N$, as well as the path they have travelled, i.e., the number of routers traversed and the average total time spent in storage, $\langle \sum_i T_s^i \rangle$. Note that signals from different frames will have a different total storage time, and so we take an average over all frames. We may then determine the average channel transmittance for each user-pairing, 
\begin{equation}\label{eq:avg_eta}
    \langle \eta_{tot} \rangle = 10^{(-\alpha L-\langle \sum_i loss_r^i \rangle)/10},
\end{equation}
where $\alpha$ is the attenuation coefficient (dB/km) of the network links, $L$ is the distance between sender and receiver, and $\langle \sum_i loss_r^i \rangle$ is the average loss over all routers in the channel, found by Eq.~\ref{eq:router_loss}. 

The simulated $N$ and $\langle \eta_{tot} \rangle$ may then be used by senders in the network to optimize their decoy-state parameters. Note that the network statistics correspond to a particular network configuration; namely, the topology, number of users, frame inter-arrival time, and routing protocol. Thus, these parameters must be known and fixed prior to a QKD session in order for user pairs to have accurate knowledge of their transmittance statistics. This is feasible in practice. For example, the network can employ traffic shaping \cite{Noormohammadpour2018} to ensure that frames from each sender arrive one at a time with inter-arrival times following the intended distribution. The remaining parameters typically do not change very frequently and their status can be updated as needed to all network users.

\section{Results and Discussion}\label{sec:results}
\begin{figure}[ht]
    \begin{center}
        \includegraphics{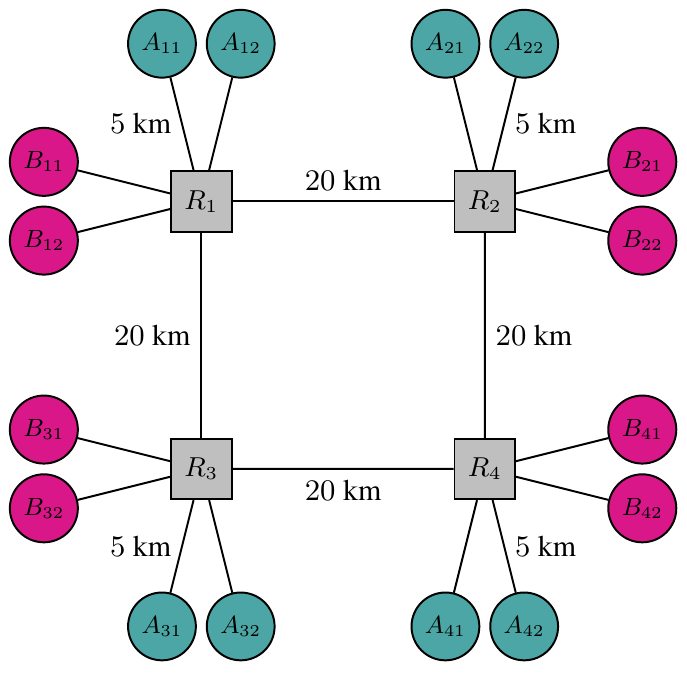}
    \end{center}       
    \caption{Sixteen-user network for simulation. Each of the four routers are connected to two Alices and two Bobs. The links are assumed to be standard single-mode optical fiber (0.2 dB/km) spanning 20 km between routers and 5 km between each user and their default router.}
    \label{fig:network-topology}
\end{figure}

In order to demonstrate the feasibility of performing QKD in a packet-switched network, we analyze the network shown in Fig.~\ref{fig:network-topology}. We choose this topology as it combines properties of star, ring, and dumbbell networks. We emphasize, however, that our approach may be used to test an arbitrary network configuration. In our simulated network, sixteen users are connected through four routers by standard single-mode fiber. In practice, each user can operate as a sender or a receiver, but we assume that users do not operate in both modes simultaneously. Thus, half of the users are designated as senders (``Alices'') and half as receivers (``Bobs''). In this section, we present the secure key rates \textit{per sent pulse} for Alice-Bob pairs separated by one, two, and three routers. We test each of the three routing protocols outlined in Sec.~\ref{subsec:protocols}. 

\subsection{No Storage During Delays}

\begin{figure*} [ht]
   \centering      
   \includegraphics{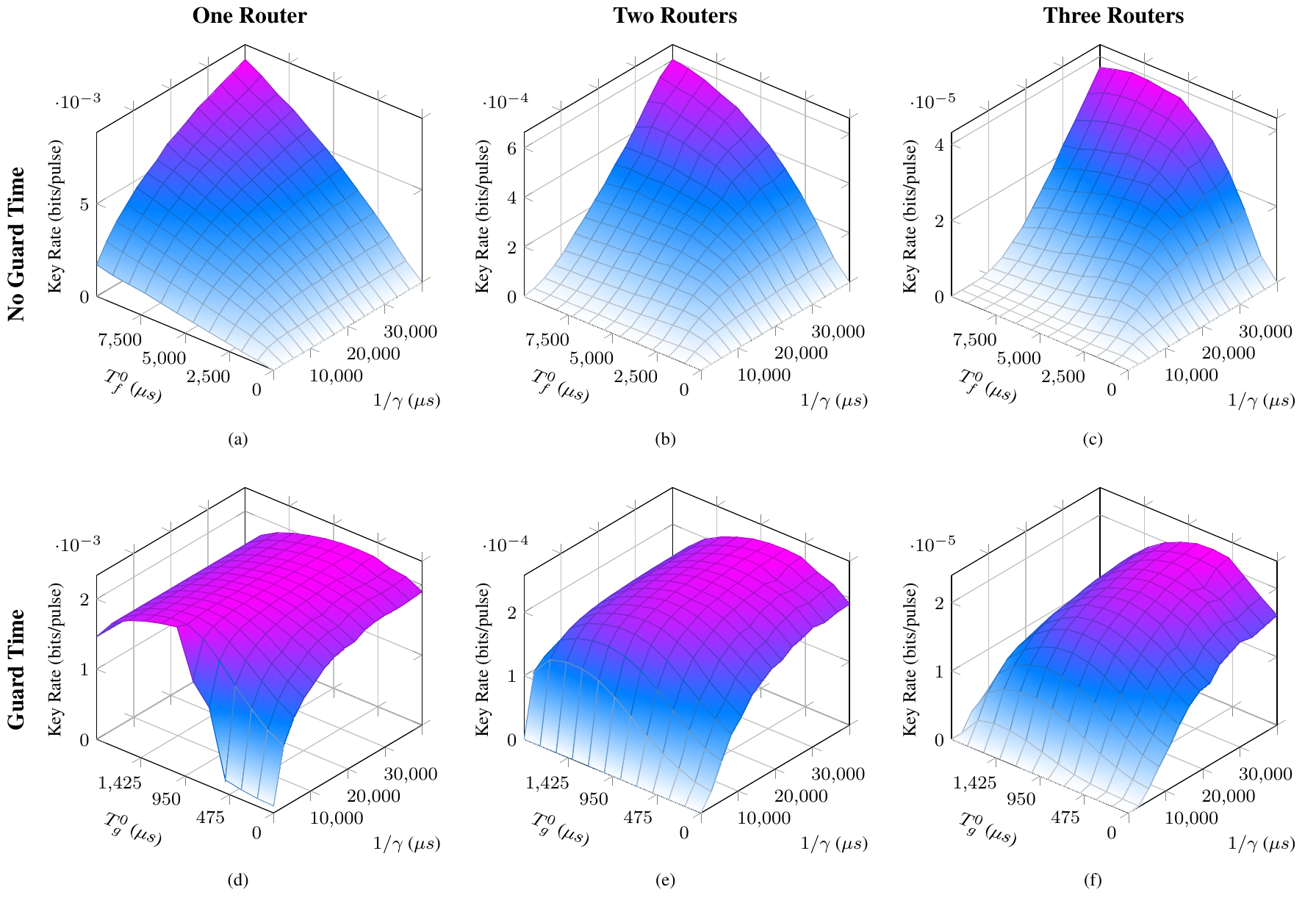}    
   \caption{Secure key rates in a network with no storage during delays. A total of 18,750 frames are generated by Alice in each user pair. The finite data size is $N\approx 10^{12}$. In plots (a)-(c), we fix the initial guard time, $T_g^0 = 0$ and vary the initial frame length, $T_f^0$ and average frame inter-arrival time, $1/\gamma$. In plots (d)-(f), we fix $T_f^0 = 2,000~\mu s$ and vary $T_g^0$ and $1/\gamma$. Columns (left to right) are for user pairs $A_{31}$ and $B_{32}$, $A_{42}$ and $B_{22}$, and $A_{22}$ and $B_{31}$ of Fig.~\ref{fig:network-topology}. Color map changes from white to purple as the key rate increases. 
   }
  \label{fig:RP1}
\end{figure*}

In Fig.~\ref{fig:RP1}, we show the key rate performance in a network with no storage during delays. We fix the number of frames sent between each user pair and examine the effects of the average frame inter-arrival time $1/\gamma$, the initial frame length $T_f^0$, and the initial guard time $T_g^0$. In this routing protocol, these parameters affect the data size, $N$, for key generation. The top and bottom rows contain the results for zero and non-zero guard times, respectively. The columns from left to right show the results for a user pair separated by one, two, and three routers.

We interpret these results as follows. Firstly, the secure key rate is expected to decrease with higher channel loss. Therefore, we observe the highest key rates for $A_{31}$ and $B_{32}$ and the lowest for $A_{22}$ and $B_{31}$. We note that due to the symmetry of the network configuration, there are negligible differences between the results of different user pairs with the same separation. For small values of $1/\gamma$, higher network traffic results in larger $d_{queue}$ leading to more pulses being discarded and thus smaller $N$. As a result, we observe a decrease in the key rate as $1/\gamma$ decreases. In Figs.~\ref{fig:RP1}(a)-\ref{fig:RP1}(c), we observe the effect of $T_f^0$. As this parameter increases, more pulses are generated. However, longer frames have a larger $d_{trans}$ which increases the time for which the server is occupied at each router and therefore increases $d_{queue}$. Thus we expect the upwards trend in the key rate to eventually stop, as is observed in Fig.~\ref{fig:RP1}(c). In Figs.~\ref{fig:RP1}(d)-\ref{fig:RP1}(f), we observe the effect of $T_g^0$ for a fixed $T_f^0$. A larger guard time means fewer pulses are discarded during delays but smaller payloads are generated. Due to this effect, we see a rise then fall in the key rate as $T_g^0$ increases. Furthermore, for a given $T_f^0$, a non-zero guard time is shown to slightly enhance the key rate. Ultimately, these results suggest that QKD can succeed in a packet-switched network even without any optical storage capacity at the routers. 

\subsection{Storage During Delays (Unlimited)}

\begin{figure*} [ht]
   \centering
   \includegraphics{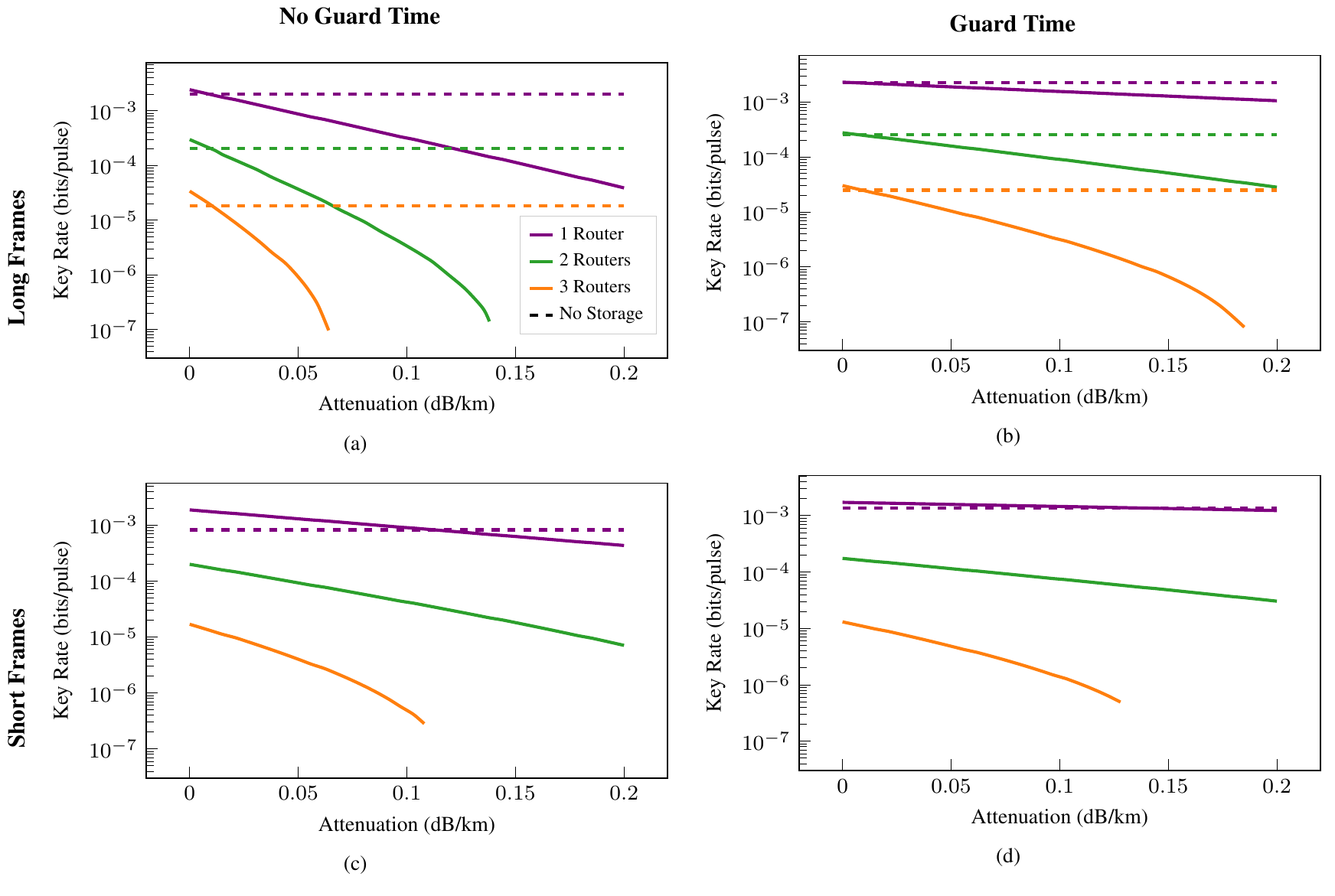}
   \caption{Secure key rates in a network with storage during delays (unlimited). In each plot, we fix the network parameters and vary the attenuation of the fiber storage lines. A total of 37,500 frames are generated by Alice in each user pair. The finite data size is $N\approx 10^{12}$. Unless displayed, the no storage routing protocol fails to produce a secure key. (a) $T_g^0 = 0$, $1/\gamma = 30,000~\mu s$, $T_f^0 = 2,000~\mu s$. (b) $T_g^0 = 800~\mu s$, $1/\gamma = 30,000~\mu s$, $T_f^0 = 2,000~\mu s$. (c) $T_g^0 = 0$, $1/\gamma = 3,000~\mu s$, $T_f^0 = 200~\mu s$. (d) $T_g^0 = 80~\mu s$, $1/\gamma = 3,000~\mu s$, $T_f^0 = 200~\mu s$.
   }
   \label{fig:RP2}
\end{figure*}

In Fig.~\ref{fig:RP2}, we show the key rate performance in a network with storage during delays, where frames have no storage time limit. We fix the number of frames sent between each user pair and examine the effect of the attenuation coefficient, $\alpha_s$, for the fiber delay lines used as storage at the routers which will determine $\langle \eta_{tot}\rangle$ for the QKD channel. The top and bottom rows consider scenarios of long and short frame lengths, respectively, where the ratio of frame length to $1/\gamma$ is fixed in each such that the average network traffic is the same. The left and right columns consider zero and non-zero guard times, respectively. For each user pair, we compare the results of this routing protocol to the no storage routing protocol under the same network parameters. 

\begin{figure}[ht]
    \begin{center}
        \subfigure[\label{fig:a_hist}]{
           \includegraphics{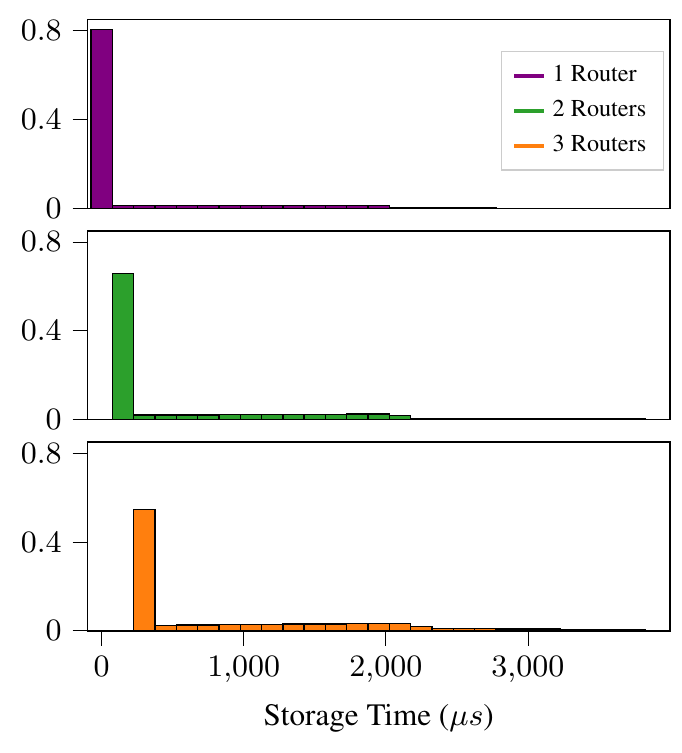}
        }
        \subfigure[\label{fig:b_hist}]{
            \includegraphics{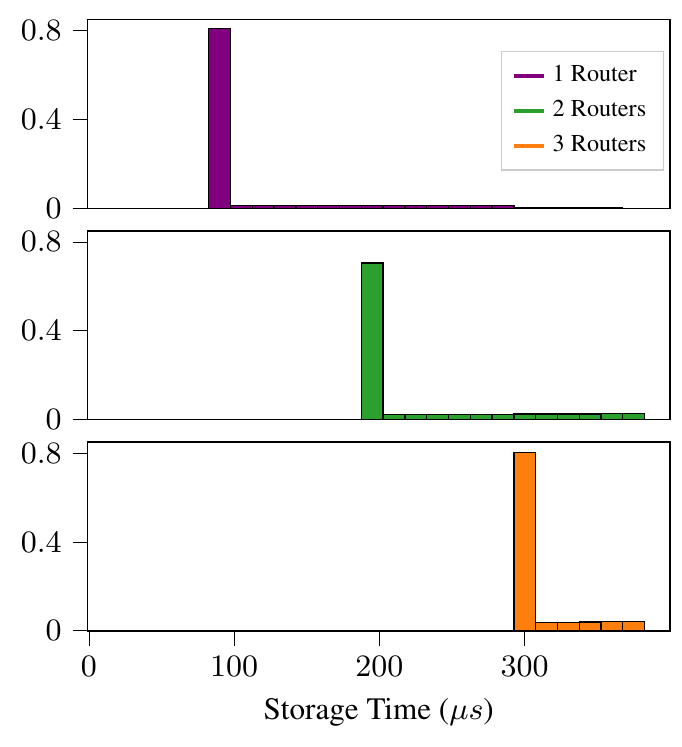}
        }                
    \end{center}
    \caption{Distribution of storage times in a network with storage during delays (unlimited). The y-axis denotes the fraction out of all frames traversing the indicated number of routers. $T_g^0 = 0$. (a) $1/\gamma = 30,000~\mu s$, $T_f^0 = 2,000~\mu s$. (b) $1/\gamma = 3,000~\mu s$, $T_f^0 = 200~\mu s$. 
    }
    \label{fig:delays_histogram}
\end{figure}

We interpret these results as follows. Firstly, the secure key rate decreases exponentially with $\alpha_s$, as expected. A non-zero guard time is again shown to enhance the key rate since it reduces the storage time of each payload, which increases $\langle \eta_{tot}\rangle$. Guard time also reduces $d_{queue}$ since it shrinks $d_{trans}$ at each router. The enhancement is more pronounced in the long frames scenario since the guard time is $\gg d_{proc}$ in this case. We observe that the short frames scenario is generally more robust to increasing $\alpha_s$, which can be attributed to smaller storage times due to a smaller $d_{trans}$. The distributions of the storage time in the long and short frames scenarios are shown in Fig.~\ref{fig:delays_histogram} for the case of zero guard time. In Figs.~\ref{fig:RP2}(a) and~\ref{fig:RP2}(b), we observe that the no storage routing protocol is generally superior when $\alpha_s > 0.01$~dB/km. We note that while attenuation coefficients as low as 0.14 dB/km have been achieved over telecom wavelengths using state-of-the-art technology~\cite{Tamura2018}, it is unrealistic to consider an attenuation much smaller than this. For a more efficient storage medium, we require long-lived quantum memories. In Figs.~\ref{fig:RP2}(c) and~\ref{fig:RP2}(d), we do not extract any secure keys with the no storage routing protocol except in the case of one router separating users. This can be explained since the frame length is on the order of $d_{proc}$, so there are zero to few non-discarded pulses from each payload. Our results suggest that, for short frames, storage during network delays is a better strategy than discarding pulses. The opposite holds true for frame lengths $\gg d_{proc}$ when we consider realistic fibers as our storage medium. This finding is important since frame lengths in a packet-switched network may have practical constraints. 

\begin{figure*} [ht]
    \centering
   \includegraphics{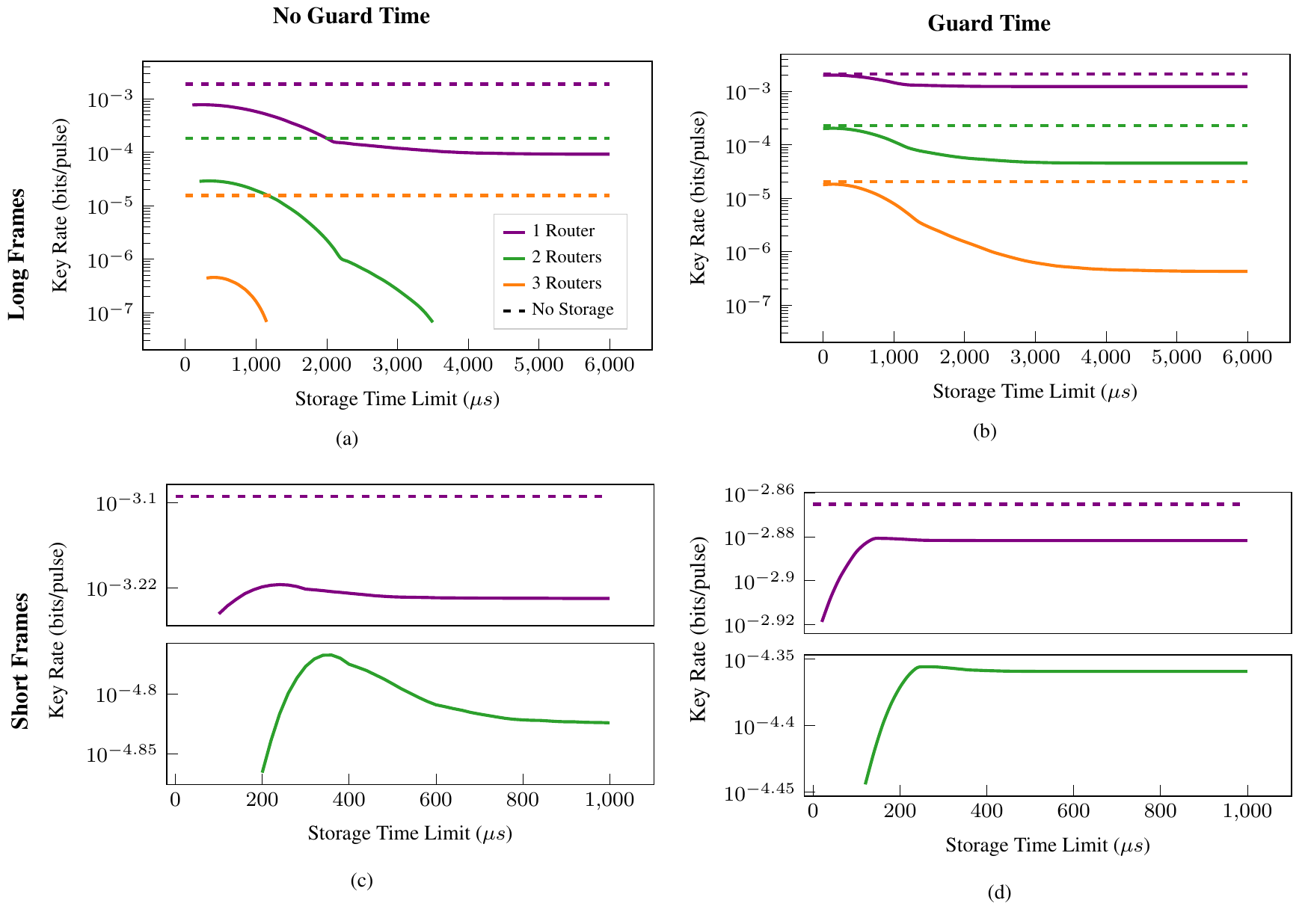}
   \caption{Secure key rates in a network with storage during delays (unlimited) and STL implemented in post-processing. In each plot, we fix the network parameters and vary the STL duration. The attenuation of the fiber storage lines is fixed at 0.16~dB/km. The network parameters are identical to Fig.~\ref{fig:RP2}.}
   \label{fig:RP2_2}
\end{figure*}

As mentioned previously, we may enforce a STL in post-processing, analogous to applying a cut-off $\langle \eta_{tot}\rangle$, in order to improve the key rate. Fig.~\ref{fig:RP2_2} shows the results for the same parameters as in Fig.~\ref{fig:RP2}, but with frames excluded from key generation if their storage time reached the STL. We consider an ultra-low-loss fiber with $\alpha_s = 0.16$~dB/km as our storage medium and examine the effect of the STL duration. It is clear that implementing a STL enhances the key rate in each scenario considered, and most significantly for frame lengths $\gg d_{proc}$. In Fig.~\ref{fig:RP2_2}(a), the optimal STL for users separated by one, two, and three routers is 200~$\mu s$, 300~$\mu s$, and 400~$\mu s$, respectively. From Fig.~\ref{fig:a_hist}, we see that these STLs preserve 82\%, 70\%, and 58\% of frames across the user pairs. In Fig.~\ref{fig:RP2_2}(b), the optimal STL is roughly 150~$\mu s$ for all user pairs and the key rates approach those of the no storage routing protocol. 

\subsection{Storage During Delays (Limited)}

\begin{figure*} [ht]
   \centering
  \includegraphics{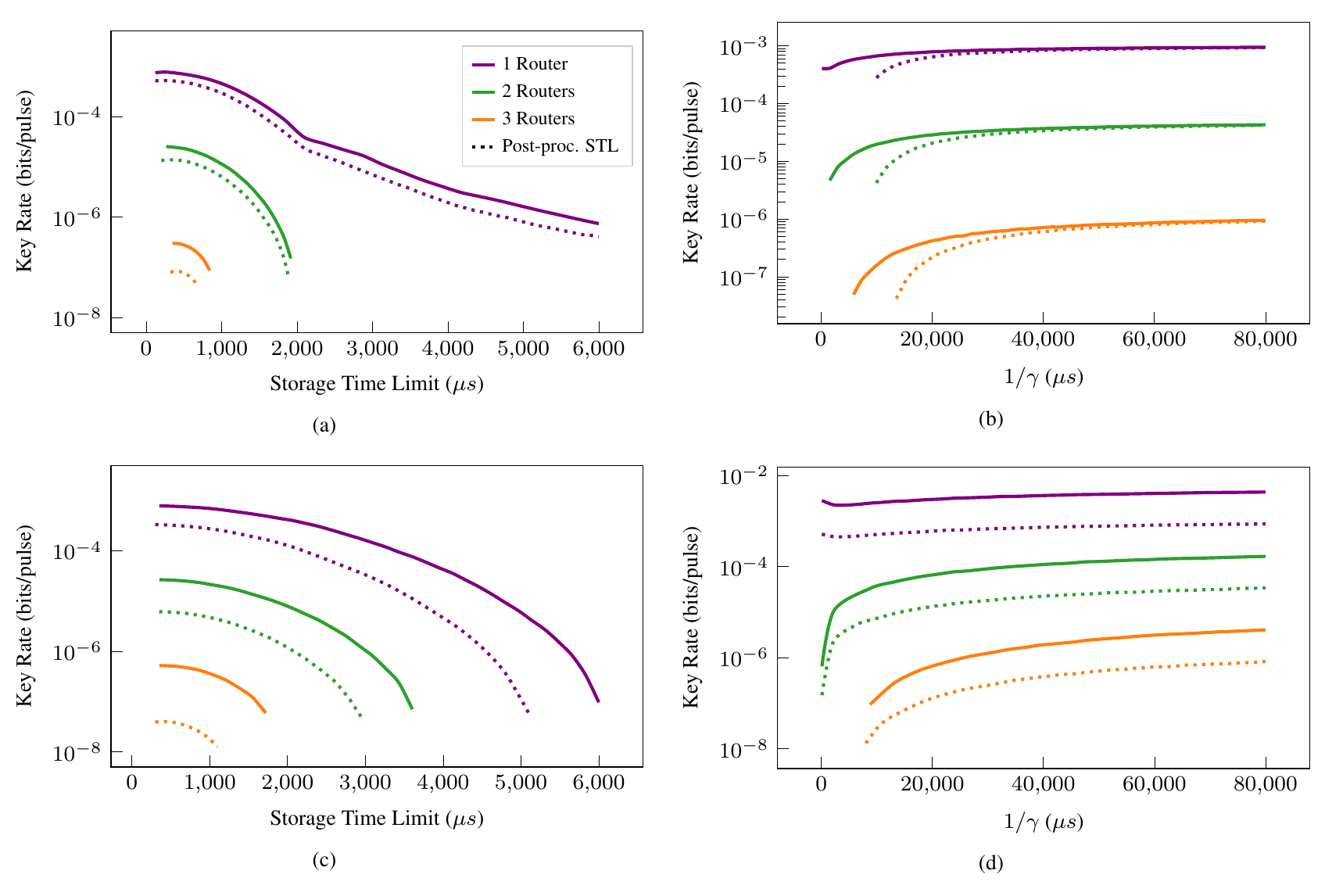}
   \caption{Secure key rates in a network with storage during delays (limited). The attenuation of the fiber storage lines is fixed at 0.16~dB/km. A total of 37,500 frames are generated by Alice in each user pair. $T_g^0 = 0$. (a) $1/\gamma = 15,000~\mu s$, $T_f^0 = 2,000~\mu s$. (b) STL~$= 320~\mu s$, $T_f^0 = 2,000~\mu s$. (c) $1/\gamma = 50,000~\mu s$, $T_f^0 = 10,000~\mu s$. (d) STL~$= 550~\mu s$, $T_f^0 = 10,000~\mu s$.
   }
   \label{fig:RP3}
\end{figure*}

In Fig.~\ref{fig:RP3}, we show the key rate performance in a network with storage during delays, where frames have a storage time limit. Once again, we fix the number of frames sent between each user pair and consider $\alpha_s = 0.16$~dB/km. We examine the effect of the STL duration under various network parameters and in each case we compare the results with the unlimited storage routing protocol where a STL is implemented in post-processing. Note that for the network parameters in the previous subsection, the two methods for implementing a cut-off transmittance produce very similar results. Here we show scenarios in which discarding frames en-route provides a significant advantage due to its mitigation of network congestion. 

\section{Outlook and Conclusions}
In this work, we have developed a framework for key rate optimization in a packet-switched network and assessed QKD performance in relation to several network parameters such as frame length, guard time, frame inter-arrival time, and storage efficiency. Notably, we found that practical secure key rates can be achieved without any optical storage capacity in the network and that guard time can generally be used to mitigate the effects of network delays. We also found that the transmittance threshold strategy used in free-space QKD can be applied in a packet-switched network to significantly enhance the key rate by limiting the permissible storage time of frames. We believe our results pave the way for future exploration of quantum applications in a packet-switched network.

Future areas of investigation may include examining more complex network topologies and perhaps a topology deployed in the field. Given that our simulation tool can accommodate arbitrary network configurations, hardware specifications, and traffic models, it can be used to establish a performance benchmark for real-world systems. The simulation tool, which we aim to make publicly available in the near future, can also be extended to examine the performance of other quantum communication tasks besides QKD such as entanglement distribution. An interesting question to address is how QKD in a packet-switched network compares to a circuit-switched network. While we have a general idea of when packet switching outperforms circuit switching based on classical networks, determining specific conditions for this advantage in a quantum network may be useful. Lastly, future work may consider the security of QKD protocols other than BB84, such as protocols where all signals sent by Alice are required to be measured by Bob. Such protocols may require us to re-evaluate the security at the routers in a packet-switched network.

\bibliographystyle{unsrt}

\appendix
\section{Storage vs. No Storage}\label{sec:nostorage_vs_storage}
\begin{figure} [h] 
    \begin{center}
        \subfigure[\label{fig:no_storage}]{\includegraphics{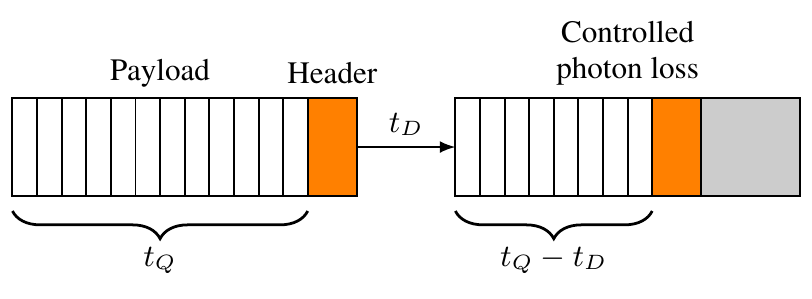}}
        \subfigure[\label{fig:storage}]{\includegraphics{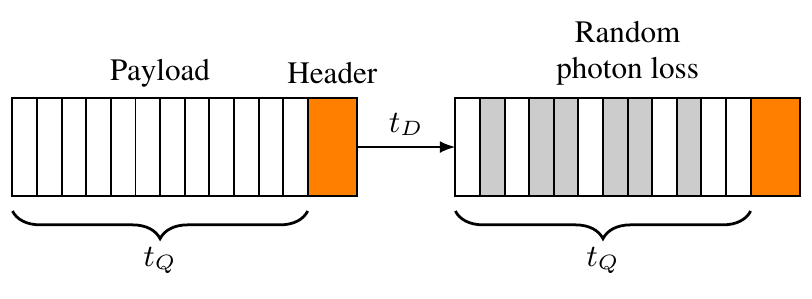}}
    \end{center}
    \caption{Schematic of a hybrid frame before and after passing a router in a packet-switched network. The duration of the quantum payload is denoted by $t_Q$. The delay time, $t_D$, is given by $d_{proc} + d_{queue}$. (a) No storage. (b) Storage.}
    \label{fig:RP1vsRP2}
\end{figure}

In this Appendix, we discuss and compare the effects of using storage versus no storage in a network routing protocol. Fig.~\ref{fig:RP1vsRP2} depicts the photon losses experienced by a payload in a router implementing these two types of strategies. We may describe the transmittance of a fiber delay line in a storage routing protocol as 
\begin{equation}
    \eta_s = 10^{-\alpha_s t_D v_g \alpha_s / 10},
\end{equation} where $t_D = d_{proc} + d_{queue}$. Thus, the number of transmitted pulses in a quantum payload of duration $t_Q$ in a storage routing protocol is given by $\eta_s R_t t_Q$. Similarly, the number of transmitted pulses in the no storage routing protocol is given by $R_t (t_Q - t_D)$. Let us assume that equal proportions of random and controlleddeterministic photon loss have the same effect on the key rate (valid so long as the detector noise is low compared to the signal). Then we expect a storage routing protocol to be favorable in the case where 
\begin{equation}
    \eta_s t_Q > t_Q - t_D.
\end{equation} Note that this comparison assumes the same network delays in each type of routing protocol. However, the no storage routing protocol introduces smaller $d_{trans}$ since frames will shrink by $t_D$ at each router they encounter. This effect leads to smaller $d_{queue}$ in the no storage routing protocol. The use of guard time in the storage routing protocol will also introduce smaller $d_{trans}$, however it is reduced by at most $T_g$.

\section{Key Rate Analysis}\label{sec:r_calc}
In this Appendix, we explain how the secure key rate is calculated and describe the optimization process. We follow the notation of Ref.~\cite{lim2014concise}. After completing the protocol described in Sec.~\ref{subsec:QKDprotocol}, Alice and Bob may distill a secure key of length
\begin{multline}\label{eq:l}
    \ell = \Bigg \lfloor s_{X,0} + s_{X,1} - s_{X,1} h(\phi_X) \\
    - n_X f_{EC} h(e_{obs}) - 6\log_2\frac{21}{\varepsilon_{sec}} - \log_2\frac{2}{\varepsilon_{cor}} \Bigg \rfloor. 
\end{multline}
Here, $s_{X,0}$ and $s_{X,1}$ are the lower bounds on the number of bits generated from zero- and single-photon pulses, respectively. The term $s_{X,1} h(\phi_X)$ is the number of bits consumed during privacy amplification, where $\phi_X$ is the upper bound on the phase error rate associated with the single-photon events and $h(x) \coloneqq -x \log_2 x - (1-x) \log_2 (1-x)$ is the binary entropy function. The term $n_X f_{EC} h(e_{obs})$ describes the bits consumed by the classical error correction algorithm \cite{Brassard1994} with efficiency $f_{EC}$, where $n_X$ is the number of detection events in basis $X$. The post-processing stage ensures that Alice's and Bob's keys are identical expect with small probability $\varepsilon_{cor}$ and secret except with small probability $\varepsilon_{sec}$. Table~\ref{tab:qkd_params} summarizes the parameter values used in the key rate analysis.  

\begin{table}[ht]
    \centering    
    {\renewcommand{\arraystretch}{1.3} 
    \caption{Key rate analysis parameters.}
    \label{tab:qkd_params}
    \begin{tabular}{ |c|c| } 
     \hline
     \textbf{Parameter} & \textbf{Value} \\
     \hline
     $f_{EC}$  & 1.16  \\ 
     \hline
     $\varepsilon_{cor}$ & $10^{-15}$  \\ 
     \hline
     $\varepsilon_{sec}$ & $10^{-10}$  \\
     \hline
     $p_{dc}$ & $2 \times 10^{-7}$  \\ 
     \hline
     $\eta_{Bob}$ & $0.15$  \\ 
     \hline
     $e_{mis}$ & $0.005$  \\
     \hline
     $\mu_3$ & $0.0002$  \\
     \hline
    \end{tabular}}
\end{table}

Next, we show how to estimate $s_{X,0}$, $s_{X,1}$, and $\phi_X$. The number of zero-photon events satisfies \cite{lim2014concise} 
\begin{equation}\label{eq:s_X,0}
    s_{X,0} \geqslant \tau_0 \frac{\mu_2 n_{X, \mu_3}^- - \mu_3 n_{X, \mu_2}^+}{\mu_2 - \mu_3},
\end{equation}
where $\tau_n \coloneqq \sum_{k \in \mathcal{K}} e^{-k} k^n p_k/ n!$ is the probability that Alice sends a $n$-photon state, and 
\begin{equation}\label{eq:n_X,k,pm}
    n_{X,k}^{\pm} \coloneqq \frac{e^k}{p_k} \Bigg [n_{X,k} \pm \sqrt{\frac{n_X}{2}\ln\frac{21}{\varepsilon_{sec}}} \Bigg]
\end{equation}
is the number of detection events in basis $X$ for pulses of intensity $k$ when considering the finite sample size. Here, $n_X = \sum_{k \in \mathcal{K}} n_{X,k}$. The detection numbers are given by 
\begin{equation}\label{eq:n_X,k}
    n_{X,k} = N q_x^2 p_k (1 - (1 - 2p_{dc}) e^{-\eta_{tot} \eta_{Bob} k}),
\end{equation} 
where $\eta_{tot}=10^{(-0.2L-\sum_i loss_r^i)/10}$ denotes the overall transmittance of the channel including the fiber links of distance $L$ (km) and the routers between Alice and Bob. Pulses belonging to different frames will have a different $\eta_{tot}$, so we take $\eta_{tot} \rightarrow \langle \eta_{tot} \rangle$ by averaging over all frames contributing to the key. Bob uses an active measurement setup with two single-photon detectors (InGaAs APDs) each with a dark count probability $p_{dc}$. The overall efficiency of Bob's measurement is given by $\eta_{Bob}$.

The number of single-photon events satisfies 
\begin{equation}\label{eq:s_X,1}
    s_{X,1} \geqslant \frac{\tau_1\mu_1 [n_{X,\mu_2}^- - n_{X,\mu_3}^+ - \frac{\mu_2^2-\mu_3^2}{\mu_1^2}(n_{X,\mu_1}^+ - \frac{s_{X,0}}{\tau_0})]}{\mu_1(\mu_2 - \mu_3) - \mu_2^2 + \mu_3^2}.   
\end{equation} 
The number of zero- and single-photon events in basis $Z$, $s_{Z,0}$ and $s_{Z,1}$, respectively, may be calculated using the same expressions by replacing $X\rightarrow Z$.

The phase error rate of the single-photon events in basis $X$ is estimated as
\begin{equation}\label{eq:phi_X}
    \phi_X \leqslant \frac{v_{Z,1}}{s_{Z,1}} + \gamma \Big (\varepsilon_{sec}, \frac{v_{Z,1}}{s_{Z,1}}, s_{Z,1}, s_{X,1} \Big ),
\end{equation}
where $\gamma(\cdot)$ is the estimation uncertainty given by 
\begin{equation}
    \gamma(a,b,c,d) \coloneqq \sqrt{\frac{(c+d)(1-b)b}{cd \log 2} \log_2\Big(\frac{c+d}{cd(1-b)b} \frac{21^2}{a^2}\Big)}
\end{equation} 
and $v_{Z,1}$ is the number of bit errors in the single-photon events in basis $Z$ estimated as 
\begin{equation}\label{eq:v_Z,1}
    v_{Z,1} \leqslant \tau_1 \frac{m_{Z,\mu_2}^+ - m_{Z,\mu_3}^-}{\mu_2 - \mu_3}.
\end{equation} Here, $m_{Z,k}^{\pm}$ is the number of bit errors in basis $Z$ for pulses of intensity $k$ when considering the finite sample size and is given by
\begin{equation}\label{eq:m_Z,k,pm}
    m_{Z,k}^{\pm} \coloneqq \frac{e^k}{p_k} \Bigg [m_{Z,k} \pm \sqrt{\frac{m_Z}{2}\ln\frac{21}{\varepsilon_{sec}}} \Bigg], 
\end{equation} where $m_Z = \sum_{k \in \mathcal{K}} m_{Z,k}$ is the number of bit errors in basis $Z$. The error numbers are given by
\begin{equation}\label{eq:m_Z,k}
    m_{Z,k} = N (1-q_x)^2 p_k (p_{dc} + e_{mis} (1 - e^{-\eta_{tot} k})),
\end{equation} where $e_{mis}$ is the error rate due to optical misalignment. 

Finally, the bit error rate in basis $X$ is calculated as 
\begin{equation}\label{eq:e_obs}
    e_{obs} = \frac{m_X}{n_X},
\end{equation} where $m_X$ is the number of bit errors in basis $X$, calculated analogously to $m_Z$.

The numerical optimization returns the parameters $\{q_x,p_{\mu_1},p_{\mu_2},\mu_1,\mu_2\}$ that maximize the secure key rate $R\coloneqq \ell/N$ for a given number, $N$, of pulses that are routed to Bob (i.e., $N$ is the difference between the number of sent pulses, $N_0$, and the number of pulses discarded in the routing process). The decoy intensities satisfy $\mu_1 > \mu_2 + \mu_3$ and $\mu_2 > \mu_3 \geqslant 0$. We fix the vacuum decoy state to be $\mu_3 = 0.0002$. The key rates presented in this paper are scaled by $N/N_0$ to show the secure key rate per sent pulse. 


\end{document}